\documentclass[12pt]{iopart}

 \usepackage{graphicx}
\begin{document}

\title[Stochastic and parameter analysis of a cancer model]{Stochastic and parameter analysis for an integrative cancer model}

\author{ Marcela V Reale$^{1,3}$, David H Margarit$^{1,2}$, Ariel F Scagliotti$^{1,2}$, Lilia M Romanelli$^{1,2}$}

\address{$^{1}$ Instituto de Ciencias - Universidad Nacional de General Sarmiento (UNGS), J. M. Gutiérrez 1150, Los Polvorines (B1613), Buenos Aires, Argentina.}
\address{$^{2}$ Consejo Nacional de Investigaciones Científicas y T\'ecnicas (CONICET), Argentina.}
\address{$^{3}$ Departamento de Ingeniería e Investigaciones Tecnológicas - Universidad Nacional de La Matanza (UNLaM), Florencio Varela 1903, San Justo (B1754), Buenos Aires, Argentina.}

\ead{mreale@campus.ungs.edu.ar}
\noindent{\it Keywords}: Cancer Immunotherapy, Cancer Cell Differentiation, Dynamical Systems, Mathematical modelling, Noise in Biological Systems, Nonlinear Dynamics

\begin{abstract}
In a previous work, we presented a model that integrates cancer cell differentiation and immunotherapy, analysing a particular therapy against cancer stem cells by cytotoxic cell vaccines. As every biological system is exposed to random fluctuations, it is important to study its stochasticity. The influence of demographic and multiplicative noise in the system is carry out on the parameters of reproduction and death in cancer cells. On the other hand, we incorporated fluctuations by adding multiplicative noise. In both cases, we analysed the dynamics for different values of the parameters involved. The final amount of cancer cells decreases for different combinations of these parameters and noise intensity is found.
\end{abstract}

%
%
%
%
%

\section{Introduction:}
Bearing in mind that cancer and the associated diseases are well studied, we will briefly describe this concept so as not to get sidetracked. For each tissue, cells have specific characteristics, stipulated times of death (apoptosis) and their rates of birth or reproduction. Cancer is a consequence of a tumour (neoplasm) in which there is uncontrolled growth of abnormal cells or when the cells lose the ability to die. In this case,  the neoplasm is said to be a malignant neoplasm. 
These cells accumulate, affecting the normal functioning of the organ that contains them and those surrounding it. Besides, they can spread to other organs generating a new malignant tumour, a phenomenon known as metastasis.
For several decades, cancer has been one of the leading causes of death worldwide. For example, in $2020$ there were around $10$ million deaths in the world caused by cancer\cite{17}. How cancer cells grow has been of great interest for years, since understand it can favour treatments against this disease. 

The usual treatments such as chemotherapy, radiation therapy, and surgery generally do not completely eliminate cancer cells. Recently, a new therapy has been used that consists of stimulating the immune system. The immune system is able to recognize tumours and eliminates many malignant cells at an early stage. But, tumours evolve to evade immune attack. Therefore, it could be possible to change or boost the immune system in order to strike cancer cells\cite{21, 22, 18,19,20}.

Developing a model for any biological system as a deterministic system, or at least not exposed to small random events, is something that can lead to a representation that is far from reality. Many times, for the modelling of biological processes, such as a process of tumour development and cancer, stochastic models are used but with the addition of different appropriate types of noise\cite{14,16}. In general, interactions between cells and microenvironmental conditions generate fluctuations both in growth and death rates, in the general populations that are interacting in the system and in the model that is developed. Examples of these can be interactions with other microorganisms, access to nutrients, temperature changes, oxygen concentration, etc.\cite{14,16,15,13}. These fluctuations may play an important role in the initial stage of the tumour, during its growing and in interaction with other type of cells, as immune cells\cite{9,8,10,11,12}. 

Based on a model developed previously\cite{1}, and taking into account the importance of simulating stochastic systems in cancer, we analysed the influence of both demographic noise (for natural birth and death rate parameters) on cancer stem cells (which we will call $S$) and in partially or fully differentiated cancer cells (which we will call $P$). This was done for different values that these parameters can take and for different noise amplitudes. In addition, as previously mentioned, we will analyse how multiplicative noise influences the entire system, taking into account analysing variations in the relevant parameters of birth and natural mortality, as well as for different noise intensities.

This paper is organized as follows: in Sec.\ref{model}, we give a description of the model is given; Sec.\ref{multi} and Sec.\ref{demo} are devoted to study the effect of the Multiplicative and Demographic noise on the system, respectively; the combined effects of cytotoxic cell vaccines and noise is studied on the Sec.\ref{impulsiva}; and, in the last section, the conclusions are drawn.

\section{\label{model} Model}
In a previous work \cite{1}, we presented an integrative model that considers the relationship between cancer stem cell $(S)$ and non-stem cancer cell $(P)$. The non-stem cancer cell include partial and total differentiated cells. This model also considers cell differentiation and their interaction with immune system cells involved in the organism natural response: natural killer $(NK)$, suppressor derived from myeloid $(M)$, dendritic $(D)$, and cytotoxic $(T)$. In order to make a more realistic model, we contemplated that there are cytotoxic that attacks $S \, (TS)$ and others that attacks $P\, (TP)$ and the same for dendritic cells ($DS$ attacks $S$ and $DP$ attacks $P$).

\begin{eqnarray}
\dot{S} & = &\alpha_S\cdot \frac{S}{S+P}+\rho_{PS}\cdot P-\rho_{SP}\cdot S-\beta_S\cdot TS\cdot \frac{S}{1+\frac{S^{1/3}}{l}}-\delta_S\cdot S \nonumber \\
& & -\mu\cdot NK \cdot  \frac{S}{1+\frac{S^{\frac{1}{3}}}{l}} \label{eq1}\\
\dot{P} &=& \alpha_P\cdot P\cdot (1-b\cdot P)+\alpha_{SP}\cdot S+2\rho_{SP}\cdot S-\rho_{PS}\cdot P \nonumber \\ & & -\beta_P\cdot  TP\cdot \frac{P}{1+\frac{P^{1/3}}{l}}-\delta_P\cdot  P -\mu\cdot  NK\cdot  \frac{P}{1+\frac{P^{1/3}}{l}}  \label{eq2}\\
\dot{TS} &=& k_{TS}\cdot \frac{DS}{DS+s_{TS}}-\delta_{TS}\cdot TS \\
\dot{TP} &=& k_{TP}\cdot \frac{DP}{DP+s_{TP}}-\delta_{TP}\cdot TP  \\
\dot{DS}&=&\gamma_{DS}\cdot S-\beta_{DS}\cdot DS\cdot TS - \delta_{DS}\cdot DS \\
\dot{DP}&=&\gamma_{DP}\cdot P-\beta_{DP}\cdot DP\cdot TP - \delta_{DP}\cdot DP \\
\dot{NK}&=&\sigma - f\cdot NK + g\cdot \frac{(S+P)^2}{(S+P)^2+h}-p\cdot NK\cdot  \frac{S+P}{1+\frac{(S+P)^{1/3}}{l}} \\
\dot{M} &=& \rho_m-\beta_m \cdot M + \alpha_m\cdot \frac{S+P}{S+P+q}
\end{eqnarray}\label{modelo_determinista}

\begin{figure}[!ht]
\centering
\includegraphics[scale=0.5]{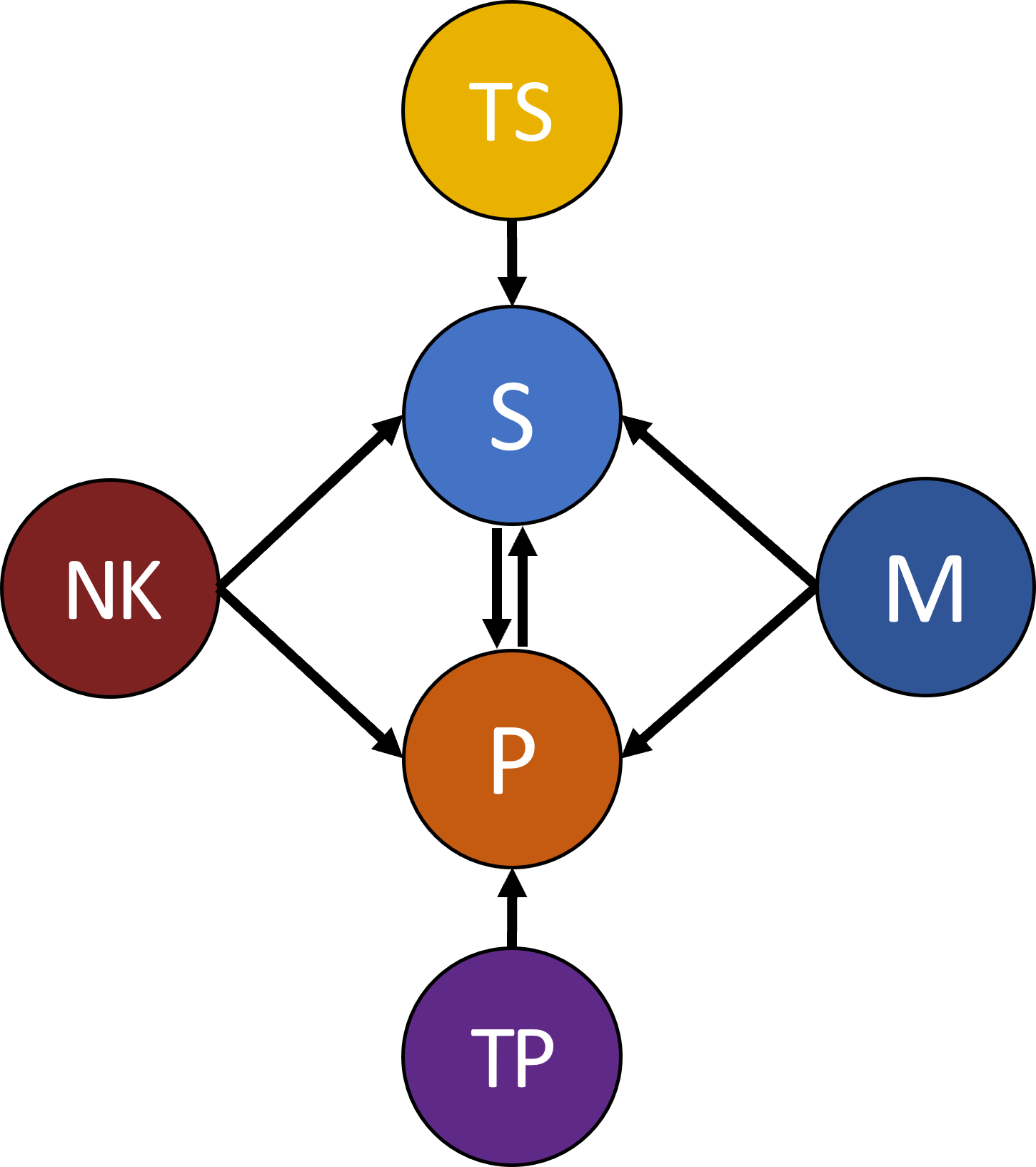}
\caption{Rough draft of the interaction between the immune system and the cancer cells described in equations \ref{eq1} and \ref{eq2}.}\label{inmunecancer}
\end{figure}

For sake of clarity, Fig. \ref{inmunecancer} shows the relations given by equations \ref{eq1} and \ref{eq2}.

The parameters range and the used values are in the Table \ref{tabla_parametros}.

\begin{table}
\caption{\label{tabla_parametros} Parameters and their ranges based in references \cite{1,2,3}}
\begin{center}
\begin{tabular}{lcr}
Parameter&Description&Value\\
$\alpha_S$ & Reproduction rate of $S$ & $0.14-0.76$ $d^{-1}$\\
$\alpha_{SP}$ & Production of $P$ through the asymmetrical division of $S$ & $0.4-0.76$ $d^{-1}$\\
$\alpha_P$ & Reproduction rate of $P$ & $0-0.8$ $d^{-1}$\\
$\rho_{PS}$ &  Plasticity & $0-0.059$ $d^{-1}$\\
$\rho_{SP}$ &  Production of $P$ through the total differentiation of $S$ & $0-0.76$ $d^{-1}$ \\
$\delta_S$ &  Death rate of $S$ due to natural processes & $0-0.25$ $d^{-1}$ \\
$\delta_{P}$ &  Death rate of $P$ due to natural processes &  $0-0.39$ $d^{-1}$\\
$l$ & Deep oh tumour accessible for immune cells &  $100$ $cell^{1/3}$\\
$\delta_{DS}$& Death rate of $DS$ due to natural processes & $0.2-0.8$ $d^{-1}$ \\
$\delta_{DP}$ & Death rate of $DP$ due to natural processes & $0.2-0.8$ $d^{-1}$ \\

$\beta_{S}$ & Death rate of cell type $S$ due to $TS$  & $6.2 \times 10^{-8}$ $\frac{1}{T_i\cdot day}$\\
$\beta_P$ & Death rate of cell type $P$ due to $TP$ & $6.2 \times 10^{-8}$ $\frac{1}{T_i\cdot day}$\\

$\mu$ & Fractional tumour cells kill rate by $NK$ cells & $3.23 \times 10^{-7}$ $\frac{cell}{day}$ \\
$k_{TS}$ & Saturated activation rate of $TS$ due to activation by $DS$  & $ 4.5 \times 10^4$ $\frac{T_i/ \mu L}{day}$\\
$k_{TP}$ & Saturated activation rate of $TP$ due to activation by $DP$ & $ 4.5 \times 10^4$ $\frac{T_i/ \mu L}{day}$\\

$s_{TS}$ & $DS$  $EC50$ for $TS$ activation rate & $6.2 \times 10^{-8}$ $\frac{mDCs}{\mu L}$\\
 $s_{TP}$ &  $DP$ $EC50$ for $TP$ activation rate & $6.2 \times 10^{-8}$ $\frac{mDCs}{\mu L}$\\

$\delta_{TS}$& Death rate of $TS$ due to natural processes & $0.02$ d$^{-1}$\\
$\delta_{TP}$ & Death rate of $TP$ due to natural processes & $0.02$ d$^{-1}$\\

$\gamma_{DS}$ & Maturation rate of $DS$ due to consumption of cancer cells & $0.0063$ $\frac{ D_i/\mu}{day/\mu L\cdot day}$\\

$\gamma_{DP}$ & Maturation rate of $DP$ due to consumption of cancer cells & $0.0063$ $\frac{ D_i/\mu}{day/\mu L\cdot day}$\\

$\beta_{DS}$ & Death rate of cell type $DS$  due to $TS$  & $6.2 \times 10^{-8}$ $\frac{1}{T_i/\mu L \cdot day}$ \\
$\beta_{DP}$ & Death rate of cell type $DP$ due to $TP$ & $6.2 \times 10^{-8}$ $\frac{1}{T_i/\mu L \cdot day}$ \\
$\sigma$ & Constant source of $NK$ cells & $ 1.4\times 10^4$ $\frac{cell}{day}$\\
$f$& Death rate of $NK$ cells & $4.12 \times 10^{-2}$ $\frac{1}{day}$\\
$g$& Maximum $NK$ cell recruitment rate &  $2.5 \times 10^{-2}$ $\frac{1}{day}$\\
$h$& Steepness coefficient of the $NK$ cell recruitment curve & $ 2.02 \times 10^7$ $cell^2$\\
$p$& $NK$ cell inactivation rate by tumour cell & $ 1 \times 10^{-7}$ $\frac{1}{ day\cdot cell}$\\
$\rho_m$& Normal $Ms$ production rate & $ 1.25 \times 10^6$ $\frac{cell}{day}$\\
$\beta_m$& $Ms$ normal death rate & $0.25 \times 10^6$ $\frac{cell}{day}$\\
$\alpha_m$& $Ms$ expansion coefficient in tumour & $1.2 \times 10^8$ $\frac{1}{cell\cdot day}$\\
$q$& Steepness coefficient of the $Ms$ production curve  & $10^{10}$ $cell$

\end{tabular}
\end{center}
\end{table}

\begin{figure}[!ht]
\centering
\includegraphics[scale=0.4]{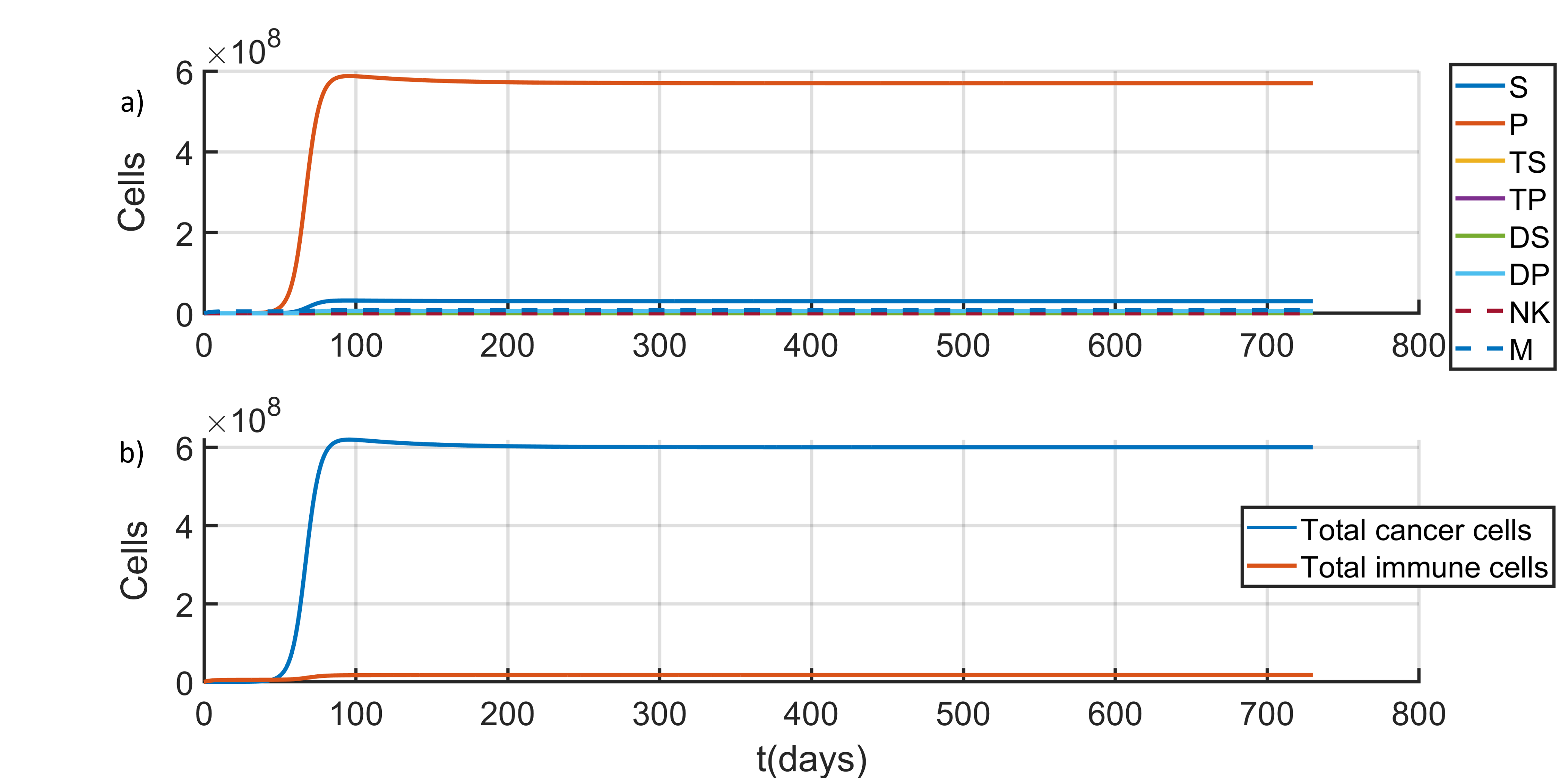}
\caption{a) Time series for all cells involved in the system. b) Time series for groups of the total cancer cells and immune cells.}\label{fig:seriestotales1}
\end{figure}

\begin{figure}[!ht]
\centering
\includegraphics[scale=0.4]{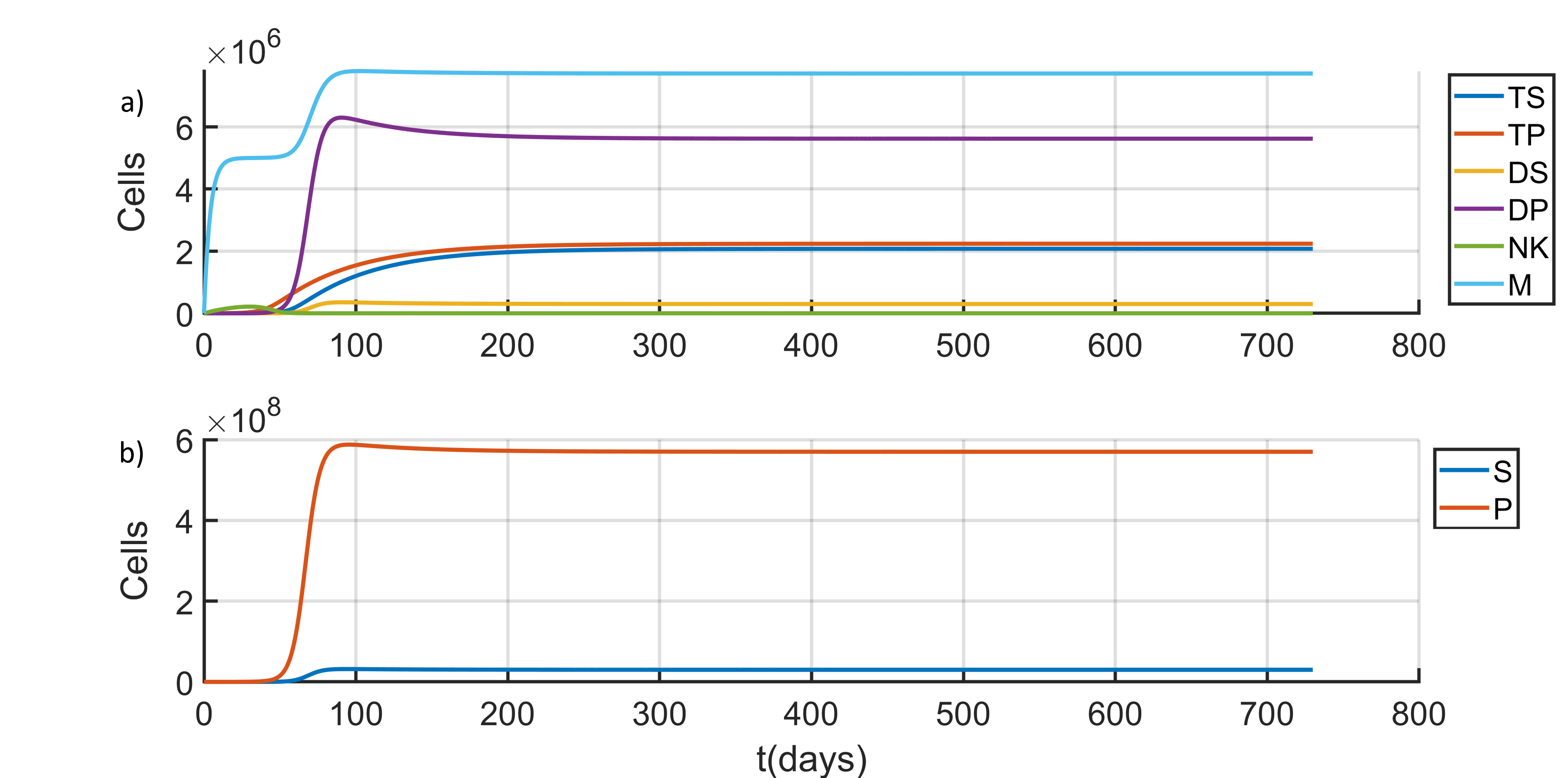}
\caption{a) Time series for each type of immune cells. b) Time series for each type of cancer cells involved in the system.}\label{fig:seriestotales2}
\end{figure}

The initial condition for all the simulations were  $S(0) = 100$, $P(0) = 2000$, $TS(0) = TP(0) = DS(0) = DP(0) = NK(0) = M(0) = 1000$. In Fig. \ref{fig:seriestotales1} and \ref{fig:seriestotales2} we can see the system evolution in time for each type and general groups (Fig. \ref{fig:seriestotales1}) and  for each type of cell, immune or cancerous, respectively (Fig. \ref{fig:seriestotales2}). For these figures, our baseline conditions for the parameters were the mean values of the ranges in Table \ref{tabla_parametros}. 

In order to quantify the results, we define  the proportion 
\begin{eqnarray*}
J = \frac{J_{final}}{J_{basal}}\nonumber
\end{eqnarray*}
where $J_{basal}$ is the mean value of total cancer cells $C$ in basal conditions and $J_{final}$ is the mean value $C$ in each situation analysed. With this definition, $ J < 1$ means that the number of tumour cells has decreased. It is important to mentioned that the results shown in this work are mean value over $10000$ realizations.

\section{\label{multi}The effect of multiplicative noise in the system}

One way to incorporated fluctuations to the system is by adding multiplicative noise to the system equations as a source of stochasticity   \cite{4,5,7}
\begin{equation}
\dot{X}i = F_i(t, X_i, X_{i+1},\ldots , X_N ) + \sqrt{D}X_i\epsilon_i(t)
\end{equation}\label{multiplicativo}
where $\dot{X}i = F_i(t, X_i, X_{i+1},\ldots , X_N )$ represents the deterministic system\ref{modelo_determinista}; $X_i$ is the population cell $i$; $\epsilon_i$ is a Gaussian white noise satisfying the conditions  $ \langle \epsilon_i(t)\rangle = 0 $ and $\langle \epsilon_i(t)\epsilon_i(t')\rangle = \delta(t-t')$; and $D$ is the multiplicative noise intensity.

\begin{figure*}[!ht]
\centering
\includegraphics[scale=0.375]{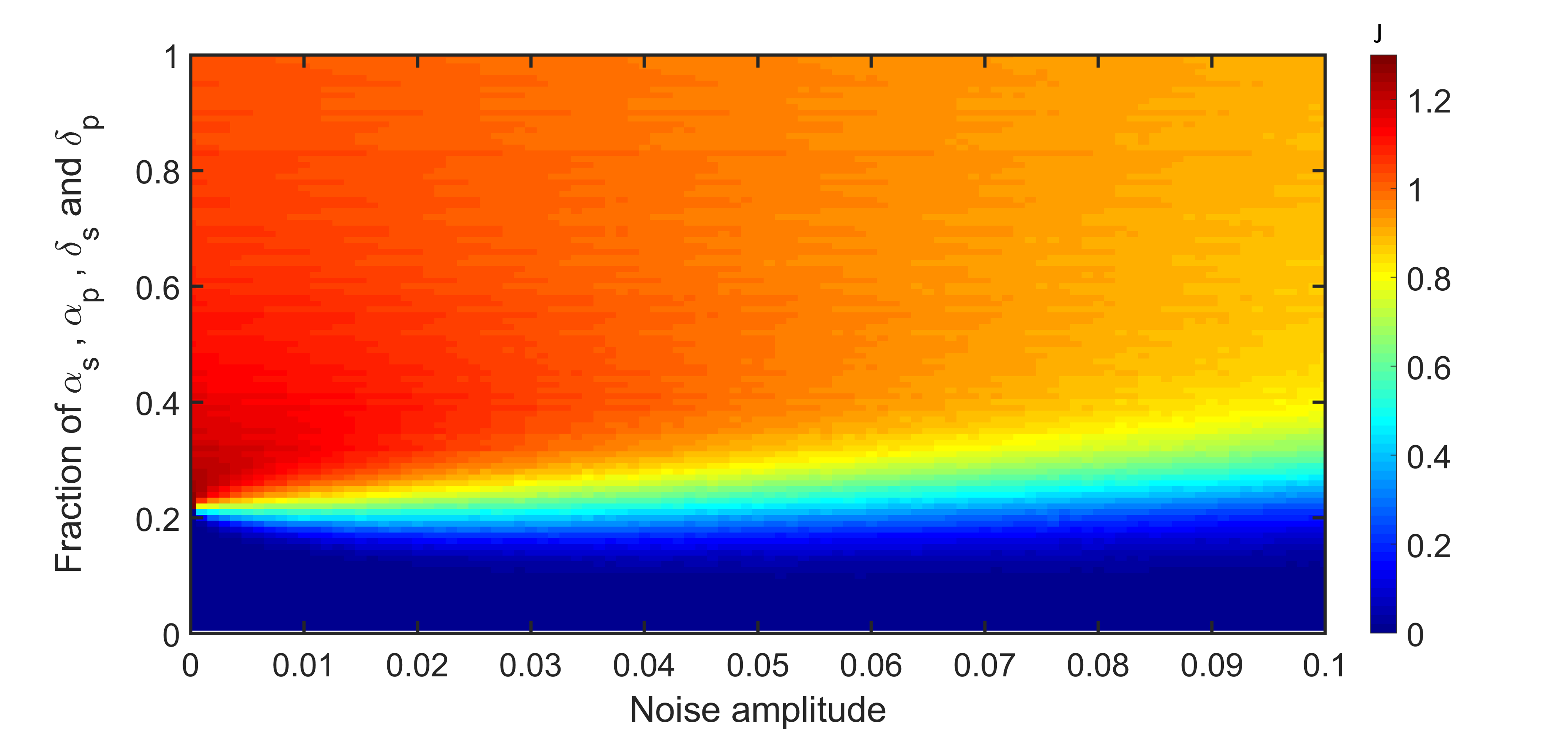}
\caption{Colour maps of $J$ for the amplitude $D$ and fraction of $\alpha_s$, $\alpha_p$, $\delta_s$ and $\delta_p$.}\label{fig_multi_j}
\end{figure*}

 In order to study the influence of the multiplicative noise, we varied the fraction of the parameters and the noise intensity $D$. On the Fig. \ref{fig_multi_j}, it is possible to observe that for different combinations of parameters, the final number of cancer cells is less than the baseline number. But we must point out that, although for small values of the fraction of the parameters the number of cancer cells is almost zero $(J\approx 0)$, it is not possible to say that by tuning these values it is feasible to completely eliminate cancer.

\section{\label{demo}The effect of demographic noise in the system}

Demographic noise or demographic stochasticity is the change in the composition of a population due to random births and deaths even though all individuals have identical birth and death rates. Usually, it is given by
\begin{eqnarray}
\alpha'_{i} & = & \alpha_{i}+ D \alpha_i\epsilon_i(t) \mbox{ for } i= S,P \nonumber\\
\delta'_{i} & = & \delta_{i}+ D \delta_i\epsilon_i(t) \mbox{ for } i=S, P \nonumber
\end{eqnarray}

where $\alpha_i$ and $\delta_i$ are the reproduction rate and the death rate, respectively; $ \langle \epsilon_i(t)\rangle = 0 $ and $\langle \epsilon_i(t)\epsilon_i(t')\rangle = \delta(t-t')$; and $D$ is the demographic noise intensity.

We analysed the influence of the demographic noise in two in different cases: first, when it is present on the parameters related to the reproduction rate of $S$  and $P$ ($\alpha_S$ and $\alpha_P$) and the death rate of $S$ and $P$ due to natural processes ($\delta_S$ and $\delta_P$); then, when it is present on $\alpha_S$ and $\delta_P$.

\begin{itemize}
\item[$\bullet$]$\alpha_S - \alpha_P - \delta_S - \delta_P$

Here we considered that all four parameters are affected with the same noise intensity. Fig.\ref{fig_demogbbdd_j} shows the values of $J$ as function of  the fraction of parameters and noise amplitude $D$. 

\begin{figure*}[!ht]
\centering
\includegraphics[scale=0.375]{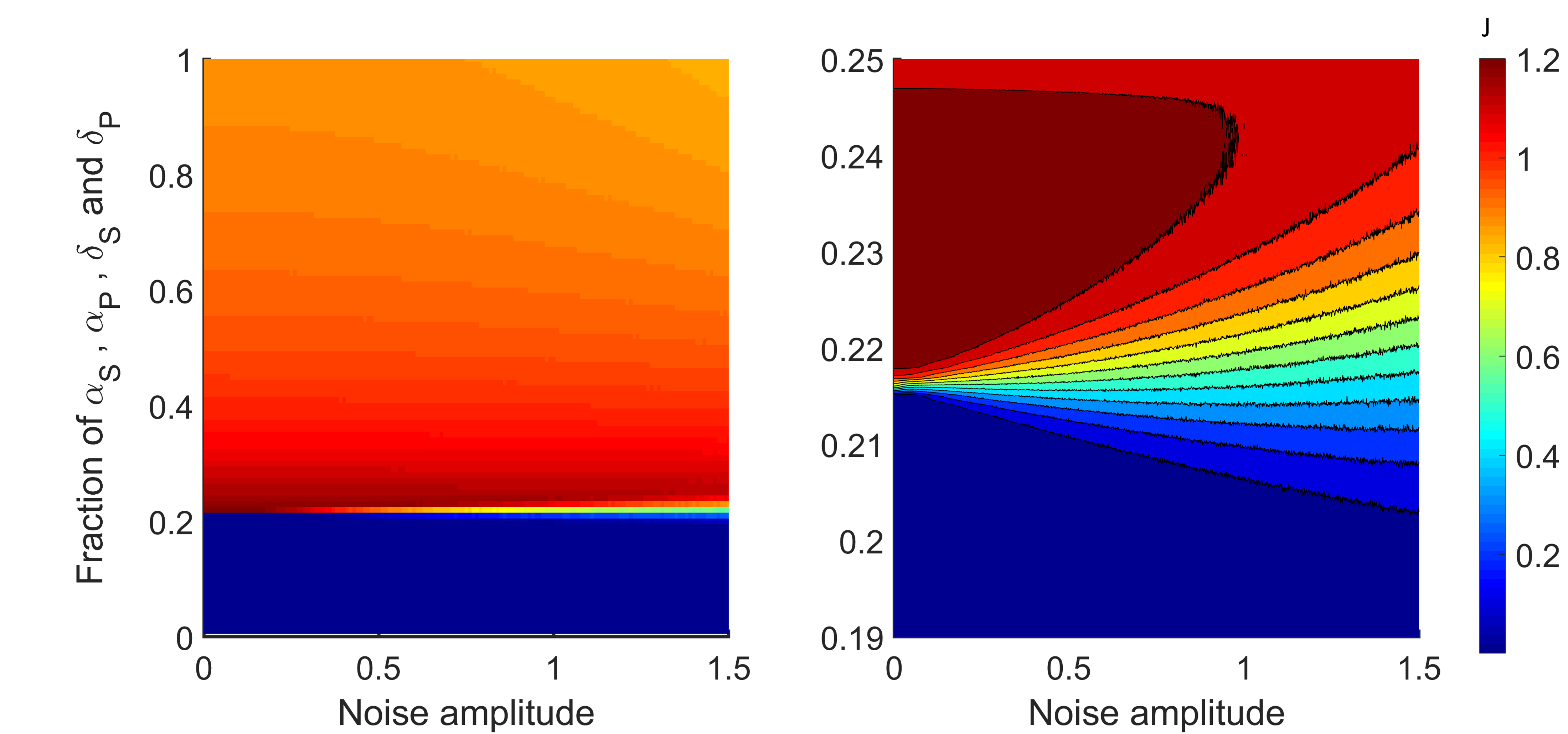}
\caption{Colour maps of $J$ for the noise amplitude $D$ in $\alpha_s$, $\alpha_p$, $\delta_s$ and $\delta_p$.}\label{fig_demogbbdd_j}
\end{figure*}

\item[$\bullet$]$\alpha_S - \delta_P$

We chose these two parameters due to that they are the most important ones: the reproduction rate of $S$, $\alpha_S$, because these cells are essential for the formation, level of aggressiveness, general growth, relapse and metastasis of a malignant tumour; the death rate of $P$, $\delta_P$, because they are bulk cells that  provides to the volume of the tumour, being some of the main responsible for angiogenesis \cite{6}.

\begin{figure*}[!ht]
\centering
\includegraphics[scale=0.375]{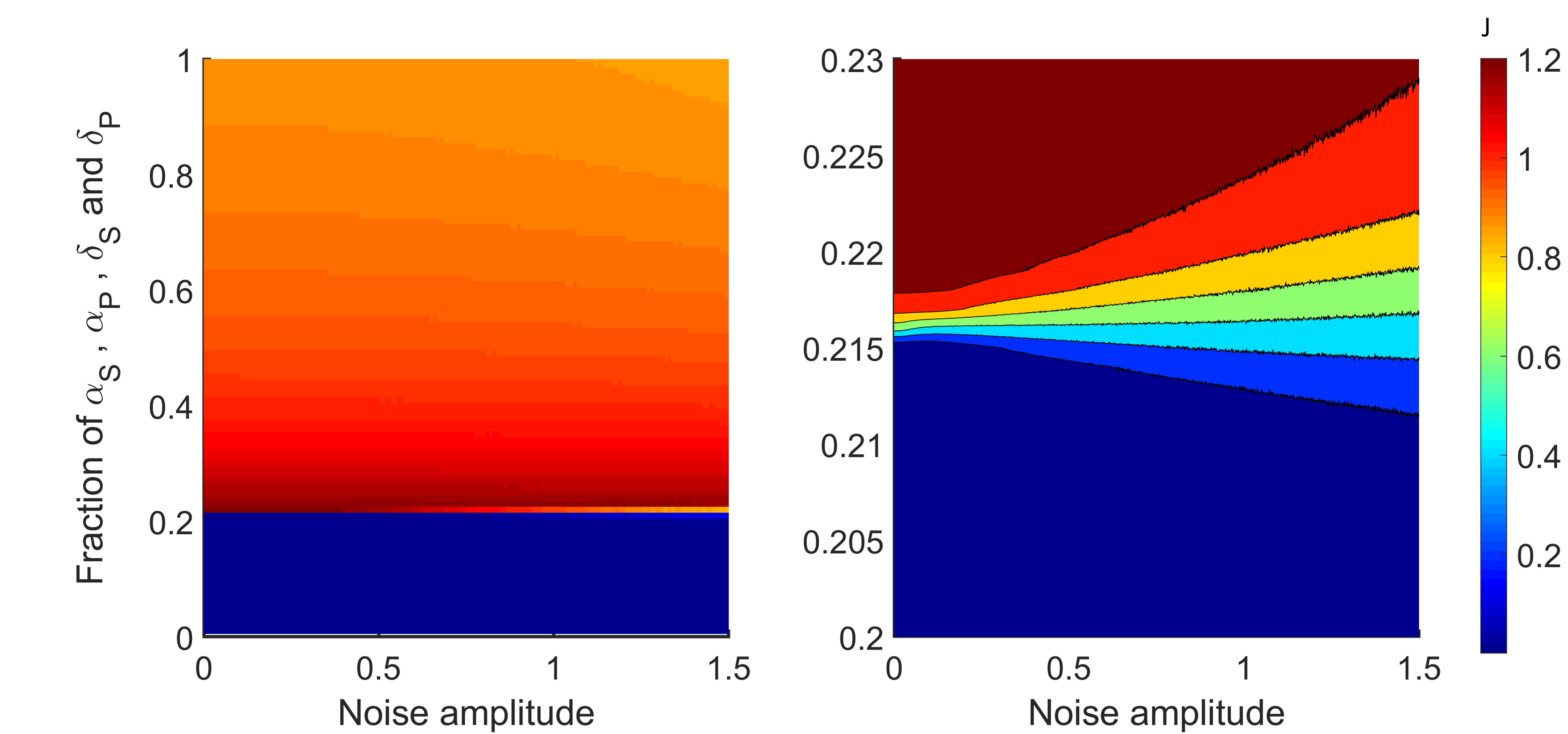}
\caption{Colour maps of $J$ for the noise amplitude $D$  in $\alpha_s$ and $\delta_p$.}\label{fig_demogbd_j}
\end{figure*}

\end{itemize}

It is worth noting that for a fixed value of the parameters and tuning the noise intensity $D$, the value $J$ varied. For some cases, the total amount of tumour cells diminished.

\section{\label{impulsiva}Study of the combined effects of cytotoxic cell vaccines and noise}

On a previous work \cite{1}, we simulated cytotoxic cell vaccines $TS$ in the system by adding on the equation  
$$ \dot{TS} = k_{TS}\cdot \frac{DS}{DS+s_{TS}}-\delta_{TS}\cdot TS $$
the function $F(t)$ is given by
$$
F\left( t\right) =\left\{
\begin{tabular}{lcc}
$A\cdot TS_{SS}$  &    for    & $ t=T\cdot K$ \\
$\left(A\cdot e^{-\beta\cdot t}+1\right)\cdot TS_{SS}$  &    for    & $ t\neq  T\cdot K$ 
\end{tabular}
\right.$$

where $T$ is the vaccination period and $K = 1, 2, 3, \ldots , n$. Thus, a treatment of period $T$, amplitude $A$ and decay $\beta$ is simulated. It is important to mention that the therapy begins on day 90.
Here, we wondered whether this treatment could be enhanced by the presence of noise. That is, if both demographic and multiplicative noise have a positive effect on the  cytotoxic cell vaccines. In other words, if it is possible to make a transition from a region where $J>1$ to another where $J<1$.

In all cases, we chose the fraction of the parameters where $J>1$ and we changed the noise intensity $D$. We setted the value of the amplitude $A$ to $1$ and simulated when the decay $\beta$ is equal to $1$ and when is equal to $2$.

\begin{figure}[!ht]
\centering
\includegraphics[scale=0.4]{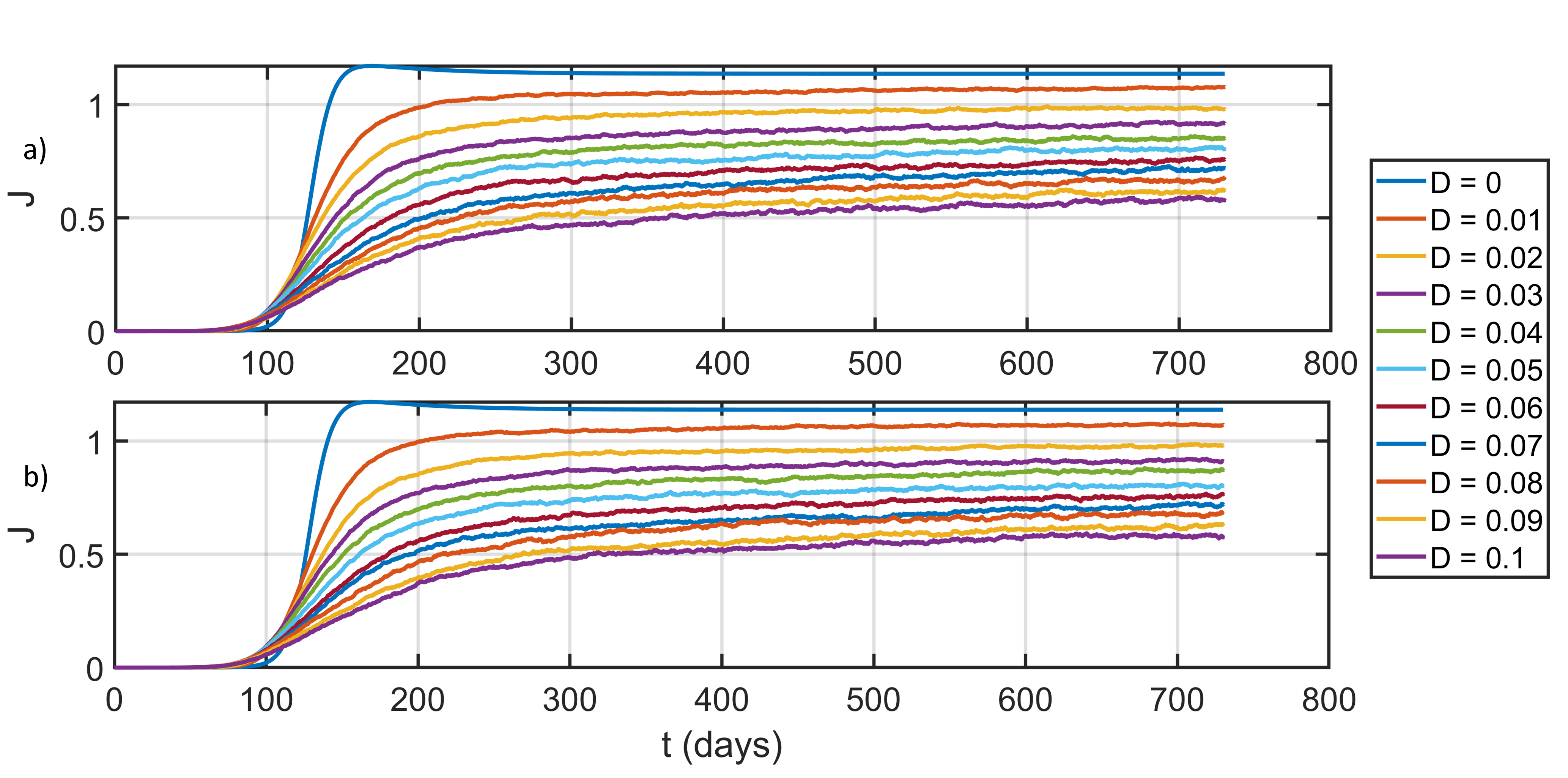}
\caption{Proportion $J$ as a function of time $t$ for different values of Multiplicative Noise intensity $D$, where $A=1$ and the fraction for the parameters is $0.3$. For a) $\beta=1$. b) $\beta=2$.}\label{fig_impulsiva_multi_1}
\end{figure}

Fig. \ref{fig_impulsiva_multi_1} shows the proportion $J$ as a function of time $t$ for different values of Multiplicative Noise intensity $D$. Here it can be seen that from a critical value $(D = 0.02)$ the total amount of cancer cells is less than at the basal conditions. It should be noted that the proportion $J$ increases with time but it does so at a lower rate than in the deterministic case $(D=0)$.

\begin{figure}[!ht]
\centering
\includegraphics[scale=0.4]{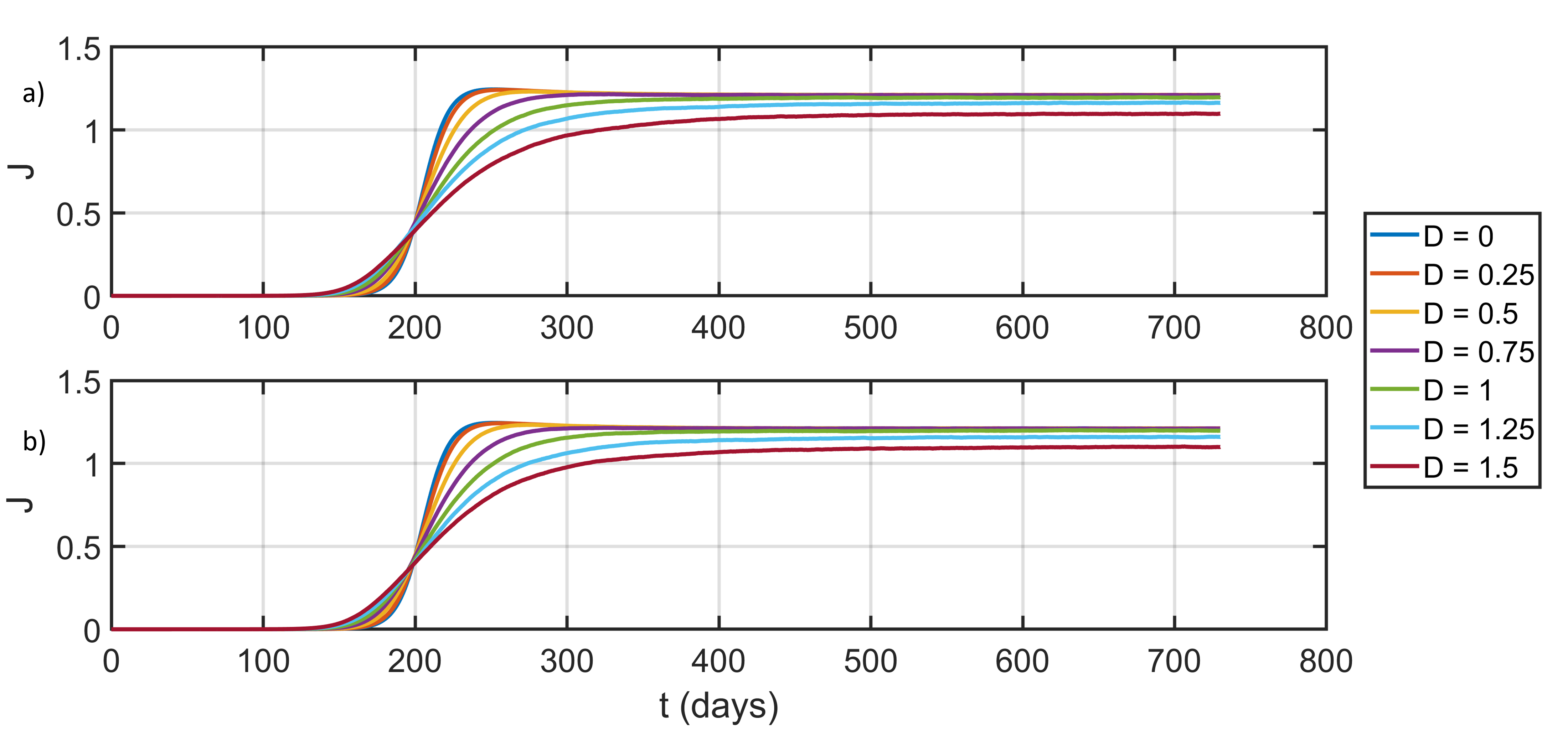}
\caption{Proportion $J$ as a function of time $t$ for different values of Demographic Noise intensity $D$ on four parameters, where $A=1$ and the fraction for the parameters is $0.24$. For a) $\beta=1$. b) $\beta=2$.}\label{fig_bbdd_impulsiva_1}
\end{figure}

\begin{figure}[!ht]
\centering
\includegraphics[scale=0.4]{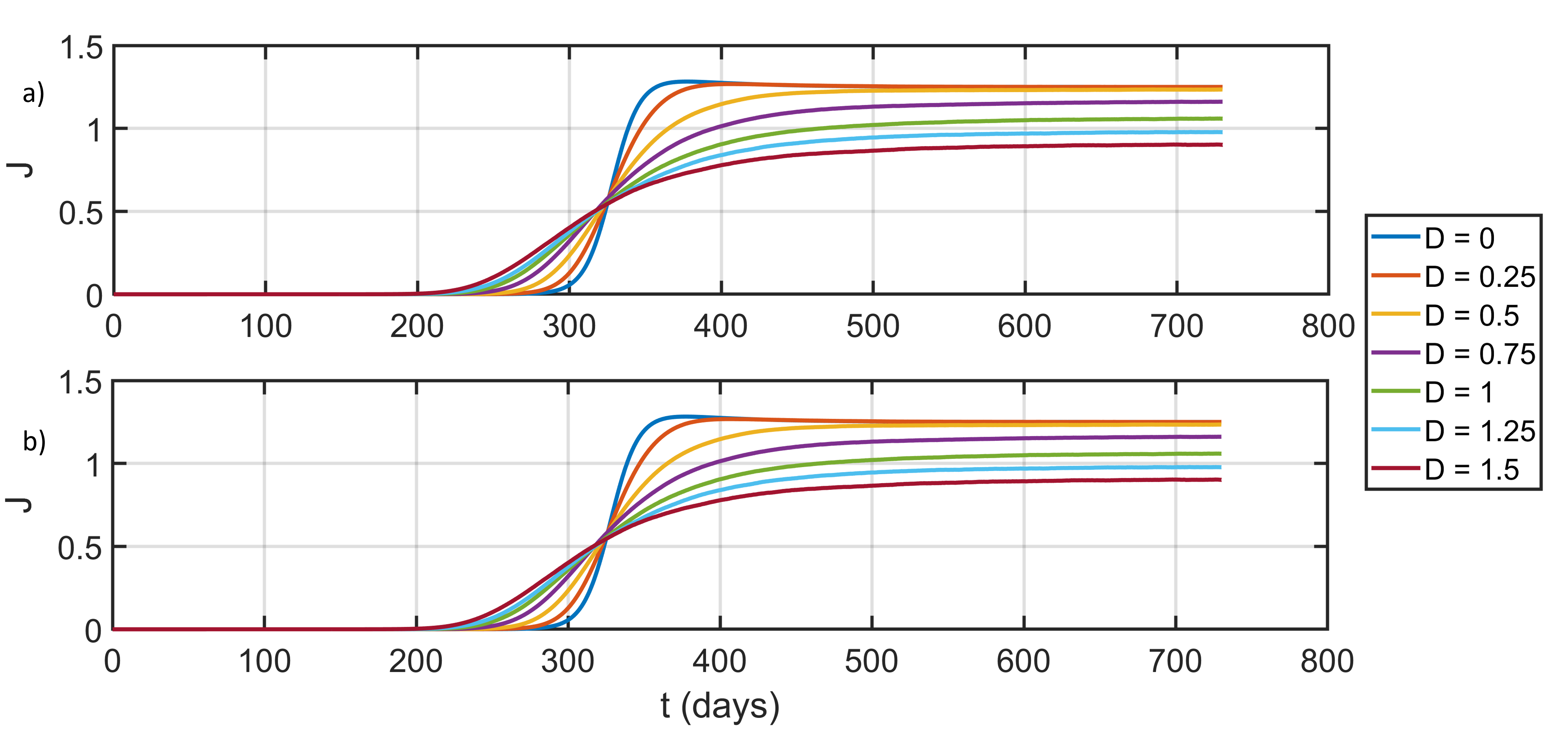}
\caption{Proportion $J$  as a function of time $t$ for different values of Demographic Noise intensity $D$ on two parameters, where $A=1$ and the fraction for the parameters is $0.22$. For a) $\beta=1$. b) $\beta=2$.}\label{fig_bd_impulsiva_1}
\end{figure}

Fig. \ref{fig_bbdd_impulsiva_1} and Fig. \ref{fig_bd_impulsiva_1} show the proportion $J$ as a function of time $t$ for different values of Demographic Noise intensity $D$ on four parameters and on two parameters, respectively. In these graphs it can be seen that as the intensity increases $D$ $J$ decreases, but never at a value lower than $1$. That is, the total amount of cancer cells was not less than the amount in basal conditions.

In general, it can be seen that the effect of multiplicative noise on the system is greater than that of demographic noise. Namely, external fluctuations have a greater influence on cancer cells.

\section{Conclusions}

Through this work, we developed an stochastic analysis of an integrative cancer model, by finding relations between relevant parameters and eventual biological fluctuations. Particularly, we studied the influence of multiplicative and demographic noise, commonly in biological systems\cite{23,24,25,26}, in an integrative cancer model for different fractions of birth and death rates. We observed that for certain combinations of fraction and noise intensity, the total amount of cancer cells is lower than the amount in basal conditions. However, the effect of demographic noise is less than the noise that comes from interactions with the environment, modelled  by addition of multiplicative noise. 

Then, we analysed how immunotherapy by cytotoxic cells is affected by these type of noises. As we can see in Fig. \ref{fig_impulsiva_multi_1}, \ref{fig_bbdd_impulsiva_1} and \ref{fig_bd_impulsiva_1}, we noticed that fluctuations of the environment have enhanced the effectiveness of treatment, but not the fluctuations of the reproduction and death rates of differentiated and non differentiated cancer cells. On the other hand, it is observed that the time in which the cancer cells reach a plateau is around 90 days (Fig. \ref{fig:seriestotales1} and \ref{fig:seriestotales2}) for the deterministic system. Unlike this, when simulating the system with noise and the vaccine treatment, the plateau is reached later between 100 and 150 days with multiplicative noise and more than 250 for demographic noise. These results, interesting to research in the future, show the positive role of noise in biological systems in combination with immunotherapy\cite{27}; since, although there is a cancer progression from day 1, this is lower than in the deterministic model.

The treatment of the immune system should be modified to achieve a greater decrease in tumour cells, in order to reach a remission phase. Therefore, this will be the objective of the future work.

\section*{Acknowledgments}

We are in debt to Santiago Doyle for helpful discussions.
This work is partially financed by PIP - CONICET N$^{\circ}$ 11220200100439CO. The authors acknowledge funding from European Union’s Horizon 2020 MSCA-RISE-2016 under grant agreement N$^{\circ}$734439 (INFERNET Project: New algorithms for inference and optimization from large-scale biological data).

\section*{References}

 \bibliography{biblio.bib}

\end{document}